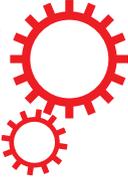

# A semi-automatic computer-aided method for surgical template design

Xiaojun Chen[1], Lu Xu[1], Yue Yang[1] & Jan Egger[2,3]

This paper presents a generalized integrated framework of semi-automatic surgical template design. Several algorithms were implemented including the mesh segmentation, offset surface generation, collision detection, ruled surface generation, etc., and a special software named TemDesigner was developed. With a simple user interface, a customized template can be semi- automatically designed according to the preoperative plan. Firstly, mesh segmentation with signed scalar of vertex is utilized to partition the inner surface from the input surface mesh based on the indicated point loop. Then, the offset surface of the inner surface is obtained through contouring the distance field of the inner surface, and segmented to generate the outer surface. Ruled surface is employed to connect inner and outer surfaces. Finally, drilling tubes are generated according to the preoperative plan through collision detection and merging. It has been applied to the template design for various kinds of surgeries, including oral implantology, cervical pedicle screw insertion, iliosacral screw insertion and osteotomy, demonstrating the efficiency, functionality and generality of our method.

Computer-assisted preoperative planning plays an important role to enhance predictability of the surgical result, in accordance with demands for accuracy, efficiency, minimal tissue damage, and even aesthetics. Aiming at transferring a preoperative plan into the actual surgical site precisely, a customized surgical template can serve as a guide to direct the implant drilling or tumor and bone resection, providing an accurate placement of the implant or prosthesis, etc.[1]. It has been widely used as an effective solution in various surgical interventions, including oral implantology, cervical or lumbar pedicle screw placement, total knee arthroplasty, treatment of dysplastic hip joint or sacroiliac joint fracture, osteotomy, etc.

Early in the 1990s, there were several reports concerning the use of manually fabricated surgical templates. Pesun and Gardner[2] described a typical technique to fabricate a template with gutta-percha for oral implant placement. Kopp *et al.*[3] designed a barium-coated template for dental implant placement with external guide wires used in conjunction with a computed tomography (CT) scan. The drawbacks of manual design and fabrication method are obvious as it is a complex process with low precision and efficiency.

Subsequently, since the beginning of the 21st century, the development of computer-aided design (CAD) and manufacturing (CAM) has brought great revolution for the design and fabrication of surgical template, and the general workflow (shown in Fig. 1) is described as follows: on the basis of the original medical images (CT, Magnetic resonance imaging (MRI), etc.), the computer-aided preoperative planning can be achieved through image processing methods including segmentation, registration, 3D reconstruction and visualization, etc., so that the ideal implant position and osteotomy trajectory can be obtained. According to this result, the surgical template can be designed, and then fabricated for clinical application using 3D printing technology. Since the template dictates the location, angle, and depth of insertion of the implant, so as to provide a link between the planning and the actual surgery by transferring the simulated plan accurately to the patient. The "in-house software" was also reported for the application of patient-specific instrument guide creation in the literature. For example, Dobbe *et al.*[4,5] developed a home-made planning software for complex long-bone deformities. With the support of this software, the interactive preoperative planning of osteotomy can be performed, and a customized cutting guide can be designed. However, the limitation is that the interobserver variation of the surgical procedure was not investigated. In addition, the software was not a general one, but just for the corrective osteotomy surgery.

[1]Institute of Biomedical Manufacturing and Life Quality Engineering, State Key Laboratory of Mechanical System and Vibration, School of Mechanical Engineering, Shanghai Jiao Tong University, Shanghai, China. [2]Faculty of Computer Science and Biomedical Engineering, Institute for Computer Graphics and Vision, Graz University of Technology, Graz, Austria. [3]BioTechMed-Graz, Austria. Correspondence and requests for materials should be addressed to X.C. (email: xiaojunchen@163.com)

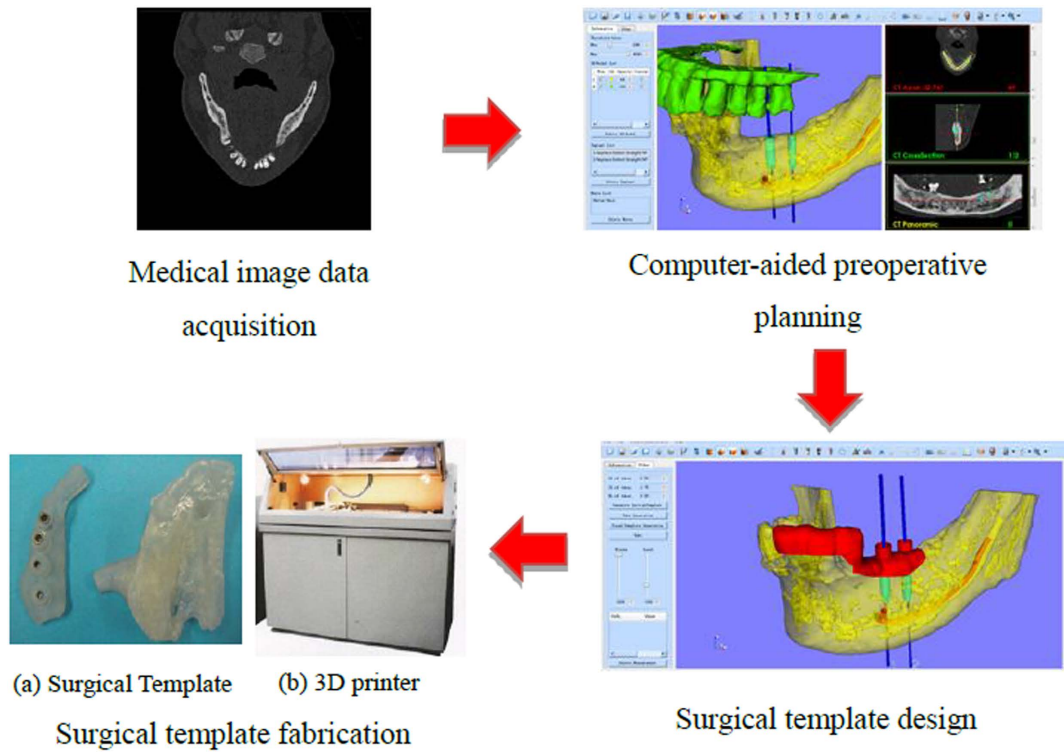

**Figure 1. General workflow of the surgical template.**

Currently, some commercially available CAD software's in industry such as Imageware (Siemens PLM Software, Germany), UG (Siemens PLM Software, Germany), Pro/E (PTC, USA), Geomagic Studio (Geomagic, USA), Paraform (Paraform, USA), CopyCAD (Delcam, UK), STTIM100 (CISIGRAPH, France), ICEM Surf (ICEM, UK), etc. have been used for the design of customized surgical templates. For example, Hu et al.[6] designed customized surgical templates through Imageware for the C2 translaminar screw insertion. Hirao et al.[7] utilized Magics RP (Materialise, Leuven, Belgium) to design a drilling template for arthrodesis of the first metatarso-phalangeal (MTP-1) joint. However, it requires high level of the engineering background to improve the efficiency of the template design, and the support from the engineers is necessary for some cases. Since the traditional CAD softwares are not dedicated for the surgical template design, the usage may be too complicated and difficult for a surgeon to learn. For example, Oka et al.[8] took several hours to design a custom-made osteotomy template for corrective osteotomy using Magics RP. Zhang et al.[9] and Chen et al.[10] reported very complicated procedures of the usage of the software of Mimics (Materialise, Leuven, Belgium) and Imageware respectively for the design of the patient-specific acetabular navigational template and iliosacral screw insertion template.

Sometimes, surgeons may need the support from professional engineers at companies for the template design. For example, Vasak[11] had to send the preoperative planning data to a certified manufacturing facility (Nobel Biocare, Kloten, Switzerland) for the design and manufacturing of a stereolithographic implantation template with appropriate guide sleeves. Stockmans[12] also reported that he received the engineering services provided by the Materialise Company (Leuven, Belgium) for the design and fabrication of the patient-specific SurgiCase® surgical guides.

Nowadays, there are also two other commercially available methods

1. Some preoperative planning software suppliers provide the services of template design and fabrication. For example, NobelGuide™ (Nobel Biocare, Gothenburg, Sweden) and SurgiGuide® (Materialise Dental, Leuven, Belgium) systems are utilized for the preoperative planning of dental implant surgery. Then, the planning data is transferred to a certified manufacturing facility for template design and manufacturing (Vasak et al.[13]). However, a relatively long delivery time is required for this kind of method. In addition, in most cases, the final products obtained from the software suppliers cannot be operated further since the surgeons are not able to participate in the process of template design.

2. There are some available software with the function of template design as well. For example, the software of Signature™ Personalized Patient Care (SPPC) (Biomet Inc., Warsaw, USA) is utilized for the design of a drilling and cutting template for total knee arthroplasty (Boonen et al.[14]). The CoDiagnostiX™ (Strau-mann, Basel, Switzerland) is used to design a drilling guide for oral implantology (Flügge et al.[15]). Although these commercial solutions allow template modification and even local design and fabrication, they are just applicable for very limited kinds of surgery. For example, the CoDiagnostiX™ and 3Shape Dental System™ (3shape A/S, Copenhagen, Denmark) are only used for the dental restoration and orthodontics, and the SPPC is for the orthopedics, etc. As for many other kinds of surgeries such as pedicle screw insertion and osteotomy, a general software for customized template design is not reported.

Therefore, semi-automatic algorithms for surgical template design were presented is this paper and then a general computer-aided design software was developed. With a simple user interface, a template can be designed and optimized through several simple interactive steps within only a few minutes. The output file is saved as the common Standard Template Library (STL) format and can be directly fabricated using 3D printing technology. Especially, the software can be utilized for various kinds of surgeries, ranging from the oral implantology as far as to the pedicle and iliosacral screw insertion.

In addition, this study involves some typical topics in the field of modeling and computer graphics including mesh segmentation, offset surface generation, Boolean operation, and ruled surface generation.

**Mesh segmentation.**    Existing mesh segmentation approaches can be roughly categorized into two groups based on the goal of segmentation[16], no matter if they are automatic, semi-automatic or interactive.

One is to segment mesh into meaningful parts, mostly volumetric, according to intuitive understanding of object components. Concavity or curvature is often utilized as key measure for the algorithms. For instance, Jagannathan and Miller[17] put forward a mesh segmentation approach using curvedness-based region growing. Au et al.[18] described an automatic mesh segmentation algorithm through locating concave creases and seams using a set of concavity-sensitive scalar fields.

The other works aim at segmenting mesh into patches under the predefined criteria or just based on the trajectory defined by the user. For example, Cohen-Steiner et al.[19] presented an approximation for the segmentation optimization problem by iterative clustering. Zhang et al.[20] proposed a feature-based patch creation algorithm for manifold mesh surfaces. Our method belongs to the second class. Here, the target region is partitioned from the input mesh to obtain the inner surface of the template according to the cutting boundary indicated by the user. In an existing method utilized by Gregory et al.[21], Wong et al.[22] Zockler et al.[23], etc., a vertex sequence in order is specified by the user, and then the mesh is segmented along the shortest path generated from the vertex sequence. However, because the path is along the edges of the triangle cells, the jaggies usually occur at the border of the segmentation result. In this study, the vertex distance scalars are utilized to partition the mesh. Original triangle cells may be cut through and new triangle cells are generated, resulting in smoother segmentation border.

**Offset surface generation.**    To obtain the offset surface of a triangulated mesh, a common method is to offset the triangles or vertices directly along the normal directions. However, the offset of triangles will lead to gaps between the adjacent triangle cells, while the offset of vertices will lead to intersection when the offset distance is larger than the minimum radius of curvature in the concave region. Several algorithms have been developed to solve this problem. Koc et al.[24] for example presented a non-uniform method for offsetting, as well as an average surface normal method to detect correct offset contours. Jun et al.[25] on the other hand, developed a curve-based method in which the curves were obtained from slicing the offset elements by the drive planes. Furthermore, in the method of Qu et al.[26], the offset vector of each vertex was calculated by the weighted sum of the adjacent facet normal. Moreover, Kim et al.[27] proposed an offset method using the multiple normal vectors of a vertex. Most of these offset methods are utilized for rapid prototype or NC milling tool path generation aiming at obtaining the precise offset result. Nevertheless, in our case, the outer surface should be as smooth as possible, only reflecting a general trend of the inner surface without details, so the precise offset is not suitable. In this study, distance field of the inner surface is established with the method of Payne and Toga[28]. Then the distance field is contoured to generate isosurface with marching cubes algorithm.

**Boolean operation and ruled surface generation.**    As a typical issue in computer graphics, Boolean operation has been studied for many years. However, even some famous commercial CAD software's have problems in Boolean operation when the models are complex. Robustness and time complexity are two major challenges. Wang[29] described an approach for approximate Boolean operations of two polygonal mesh solids with Layered Depth Images. However, the resulted mesh may be self-intersected. Feito et al.[30] presented a method for Boolean operation on triangulated solids, which is based on a straightforward data structure and the use of an octree. We, however, aim at simplicity and reduction in time complexity, Boolean operation is simplified for collision detection and merging in that only "union" is employed in our case so that the automatic identification of relevant parts according to Boolean operation type can be omitted.

As for the ruled surface, it is utilized to merge patches together, including the connection of the inner and outer surfaces, and merging relevant parts after collision detection. Fuchs et al.[31] proposed a valid method to simplify the problem of ruled surface determination to shortest-path problem in a directed graph. However, the algorithm for the solution of shortest-path is complicated, and not suitable for our case. So we proposed a label setting algorithm that is utilized for the solution of shortest-path problem.

## Results

A general framework of semi-automatic surgical template design was introduced and several algorithms including inner surface generation, outer surface generation, ruled surface, collision detection and merging were presented. On the basis of these algorithms, a software named TemDesigner (The screenshot of the software is shown in Fig. 2) was developed under the platform of Microsoft Visual Studio 2008 (Microsoft, Washington, USA). Some famous Open Source toolkits including VTK (Visualization Toolkit, an open-source, freely available software system for 3D computer graphics, image processing and visualization, http://www.vtk.org/) and Qt (a cross-platform application and UI framework, http://qt-project.org/) were involved. Several cases of customized template design for various kinds of surgeries were conducted using TemDesigner. No specific condition was required for mesh tessellation or concavity for those cases. With the manually-drawn curves indicating the target regions and relative input parameters, the templates can be generated automatically and rapidly. The results shown in below demonstrated the effectiveness and generality of our approach.

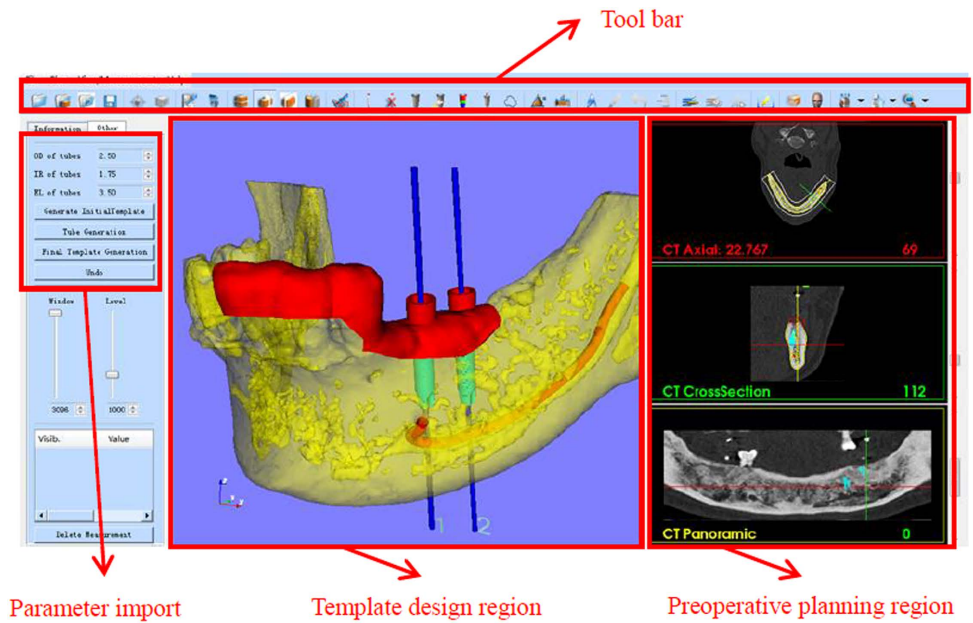

**Figure 2. A screenshot of the TemDesigner.**

### Oral implantology.

The preoperative planning for oral implantology was accomplished through the software named CAPPOIS[32] (Computer-Assisted Preoperative Planning for Oral Implant Surgery, Institute of Biomedical Manufacturing and Life Quality Engineering, Shanghai Jiao Tong University, Shanghai, China) to determine the optimal positions and orientations of implants. The surface mesh of dentition was generated based on the registration, which means superimposing the three-dimensional laser-scanned model of plaster casts of dentition onto the three-dimensional skull model reconstructed from CT images. The detailed design procedure is shown in Fig. 3. Firstly, initial control points were indicated by the user and the contour curve was generated and updated dynamically (Fig. 3(1)). After the target region was determined (Fig. 3(2)), the initial base template without drilling tubes was generated automatically (Fig. 3(3)). Then, the axes of virtual preoperative planned implants were imported, indicating the positions and orientations of the drilling tubes (Fig. 3(4)). With related parameters including inner and outer radii and length of tubes input by the user, the final tooth-supported template was generated automatically (Fig. 3(5,6)).

### Cervical pedicle screw placement.

For the pedicle screw placement, how to provide good anchoring without unexpected perforation poses a great challenge for surgeons. Intraoperative navigation using optical tracking device can be an effective method. However, the registration process is usually quite time-consuming. For each vertebra, a separate registration step is demanded, which typically spends about 15 minutes[33]. This means the operating time will be increased with the added amount of vertebras for insertion, leading to a higher risk of intraoperative infection. The use of a surgical template is a feasible solution. In our case, the surface mesh of vertebral column was reconstructed with CT data using Slicer 4.3 (a free, open source software package for visualization and medical image computing, http://www.slicer.org/). Figure 4 shows the process of template design for the cervical pedicle screw insertion. The target region was defined to cover the lamina and spinous process in a lock-and-key type for stability of positioning during the surgery. The thickness of template was set as 2.5 mm to ensure suitable strength. The orientations of drilling tubes were determined according to preoperative planned trajectory.

### Iliosacral screw insertion.

The iliosacral screw fixation has been widely used for the stabilization of unstable pelvic fractures. Customized templates for the iliosacral screw insertion can be a good option to achieve accurate screw placement, reduction of radiation exposure and surgical time compared with traditional methods of fluoroscopic detection. The CT data of the patient were imported into Slicer 4.3 to reconstruct the 3D model of pelvic girdle. The target region was designed to cover the iliac crest for fixation during surgical operation. The drilling tube was oriented through the sacro-iliac joint into sacrum. The design procedure and results are illustrated in Fig. 5.

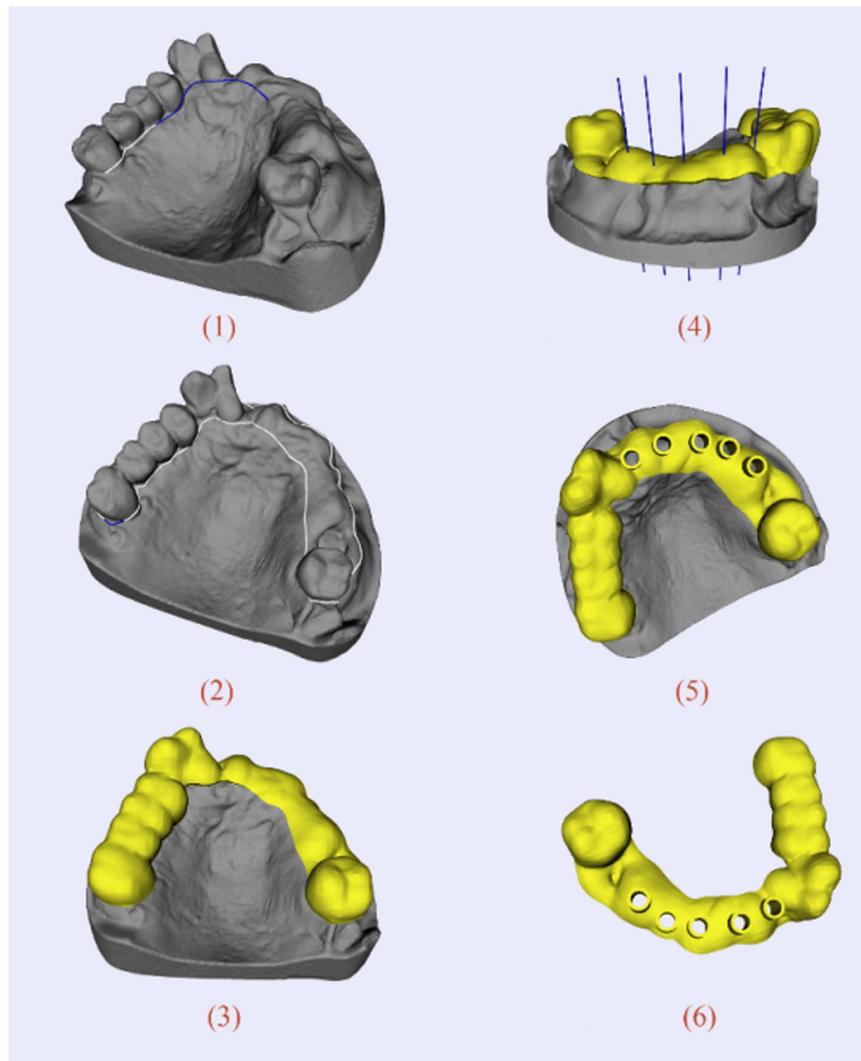

**Figure 3.  A typical template design process for oral implantology with TemDesigner:** (**1**) Import the 3D model and indicate points surrounding the target region. The curve will be generated and updated dynamically; (**2**) The target region is determined by the closed curve; (**3**) Initial template without drilling tubes is generated automatically; (**4**) Import the axes of virtual implants; (**5,6**) Final template is generated.

**Osteotomy.**  Customized templates are widely used for the treatment of cubitus varus deformity in osteotomy. Different from the templates mentioned above, there's no drilling tube on the template of osteotomy. During the surgery, the template is placed at the target region of the bone. Then, the bone can be resected along with the borderline of the template. Figure 6 shows the design procedure and the result of a template for osteotomy.

In order to evaluate the quality of the designed guides, the actual template and adjacent tissue models have been fabricated through the 3D printing technology (shown in Fig. 7). The verification result demonstrated the unique topography between the match surface of the templates and the adjacent tissues. In addition, the previous pilot study[32] proved that the fixation of the templates was unique, stable, and reliable, and the accuracy of surgical outcome can meet the clinical requirement and more clinical trials will be conducted in the future.

All the experimental results were conducted on a PC with Intel Core i5-3210 with a 2.50 GHz CPU, 6 GB memory and a 64-bit Windows 7 operating system. Table 1 shows the property of the input surface models and the corresponding computing time of each step during the procedure of the template design. In this table, time of the initial template generation is the sum of 'Inner surface segmentation', 'Offset of inner surface', 'Generation of points for outer surface segmentation', 'Outer surface segmentation' and 'Connection of inner and outer surfaces'. For all examples, the overall time for the automatic computing is less than one minute. As for the surface segmentation, the computational complexity of this algorithm is $O(N \cdot n)$, where N is the number of points of the input mesh for segmentation, and n is the number of points for segmentation. That means that the time for inner surface segmentation depends on the scale of the input surface mesh and the length of the curve of target region. For offset of inner surface, the computing time depends on the scale of the inner surface. As for the ruled surface generation, the computing time of this part depends on the sampling step, and the scale of the inner surface and offset surface. Furthermore, the user interaction time including the generation of contour curve and the related parameters setup (such as the thickness of template, the size of implant, etc.) was approximate 8–10 minutes. It was related to the type of surgery, the size and shape complexity of the 3D-reconstructed models, the user operation proficiency, etc.

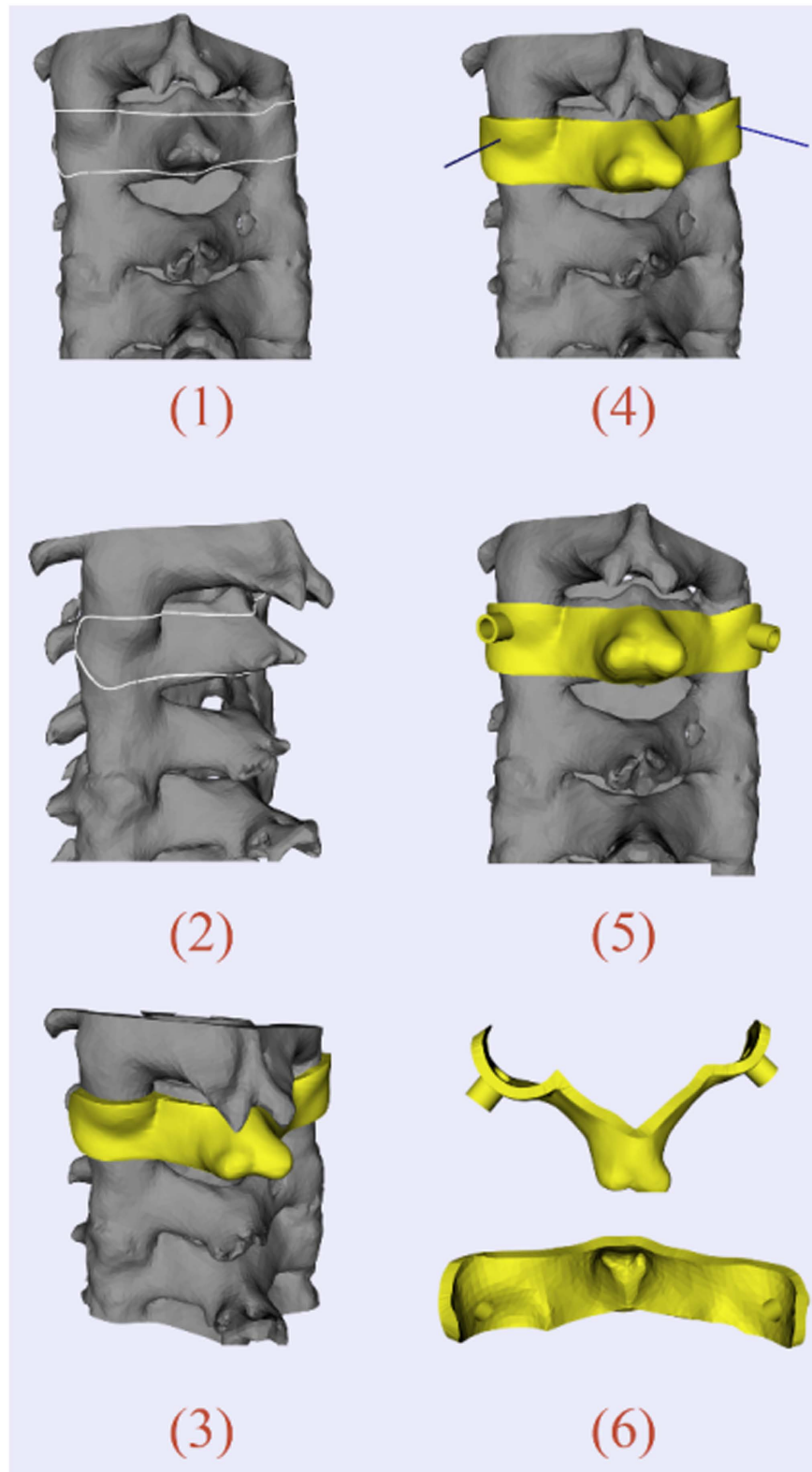

**Figure 4. Template design for cervical pedicle screw placement:** (**1,2**) Model of cervical vertebrae. White curve indicates the target region—border of the inner surface; (**3**) Initial template generation; (**4**) Blue line segments show the axes of virtual implants; (**5**) Final template positioned on surgical site; (**6**) Vertical view and inner surface of the template.

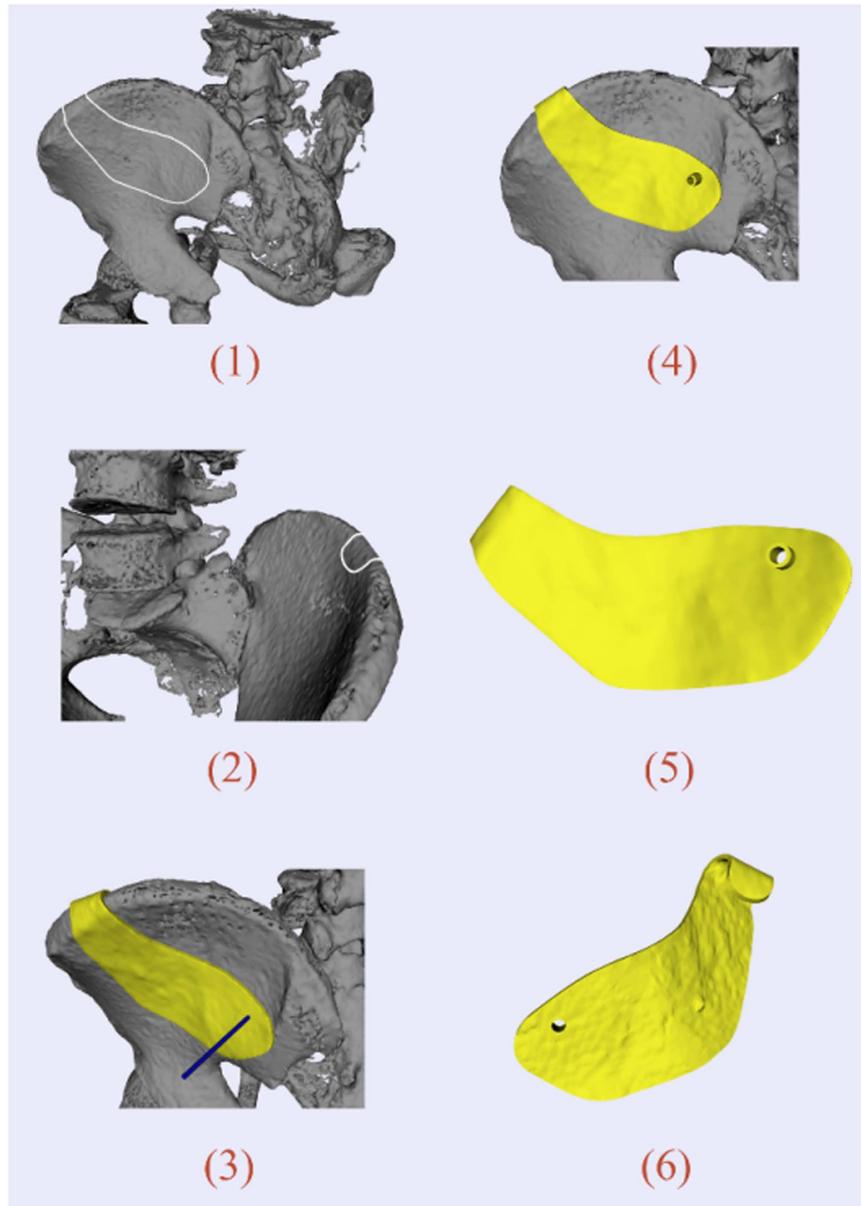

**Figure 5. Template design for iliosacral screw insertion:** (**1,2**) Model of the pelvis. White curve indicates the target region; (**3**) Initial template and axis of the drilling tube; (**4**) Final template positioned on surgical site; (**5**) Outer surface of the template; (**6**) Inner surface of the template.

## Discussion

The method for the semi-automatic template design in this study is described as follows: Firstly, a point loop is indicated by the user on the target mesh. Then, for each vertex of the mesh, a signed distance to the input point loop is calculated for contouring to achieve mesh segmentation. Hence, the inner surface is clipped from the entire mesh for exact match and stability of positioning during surgical operation. Subsequently, the distance field of the inner surface is calculated to obtain the offset surface, which will afterwards be segmented to obtain the outer surface of the template. In order to form a closed model, the ruled surface is employed to connect the inner and outer surfaces. The generation of the ruled surface is transferred to the shortest path problem in a directed graph. Finally, the Boolean operation, which is simplified to collision detection and merging, is utilized to add the drilling tubes to the template.

In conclusion, based on the above-mentioned algorithms of TemDesigner presented in the section of "Introduction", surgeons can design and modify the template efficiently with simple interactions, and then fabricate it using 3D printing technology. It can be a very effective way for template design to reduce the delivery time.

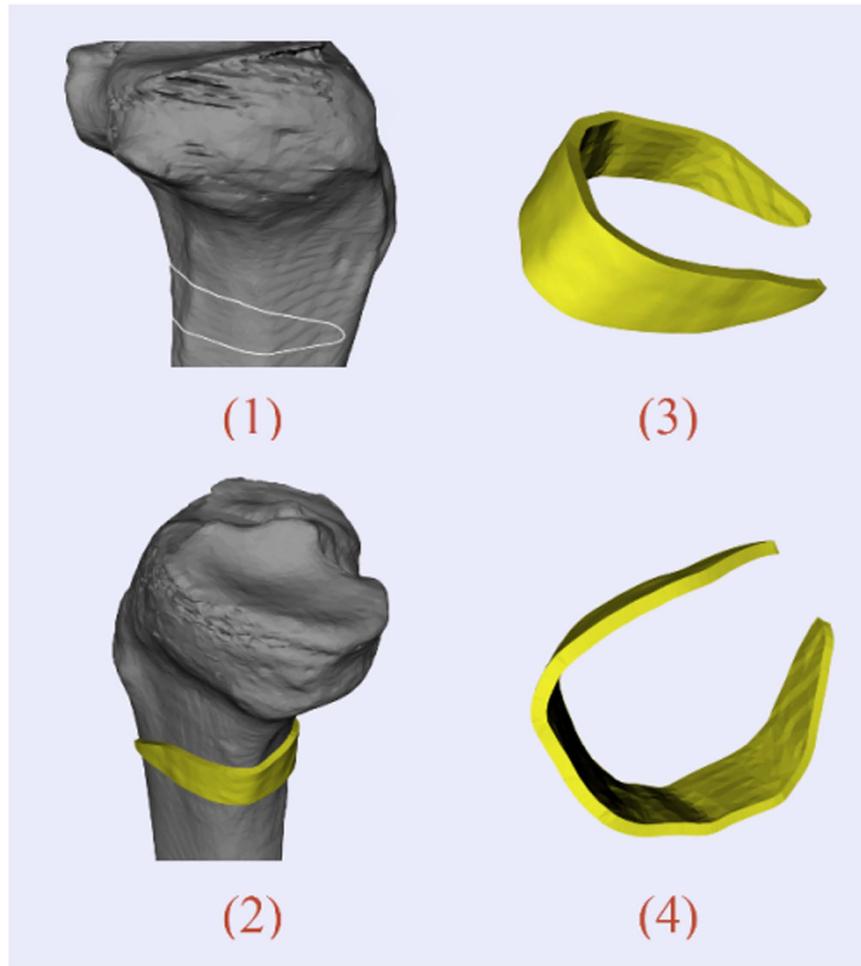

**Figure 6. Template design for osteotomy:** (**1**) Model of the bone. White curve indicates the target region; (**2**) Final template positioned on surgical site. The bone will be sectioned along the borderline of the template; (**3,4**) Different perspectives of the template.

Drilling and cutting templates for various surgeries, including oral implantology, cervical pedicle screw insertion, iliosacral screw insertion and osteotomy have been semi-automatically designed with TemDesigner, demonstrating the functionality and generality of our method.

In addition, our method using TemDesigner has been compared with the other two kinds of methods (Method 1: Using the Imageware, UG, and Magics RP together; Method 2: Using 3-matic).

The workflow of Method 1 for the template design (take oral implantology as an example) is shown in Fig. 8. Since the functions involved in these three softwares are very complicated, it requires high level of the engineering background for the user to grasp all of these softwares and accomplish the template design.

As for Method 2, the functions of combining surfaces, repairing and de-featuring, remeshing, modifying and editing, etc. are used for the template design. Although the complexity level of the usage of the 3-matic is lower than Method 1 (for example, no need of importing and exporting), the user is still required to own the engineering background knowledge of geometry design and get very familiar with the 3-matic.

With respect to TemDesigner, the user is only required to indicate some initial control points so that the contour curve is obtained and updated dynamically. Then, after inputting some related parameters, the final surgical template is generated automatically. It also means our method does not require a lot of skills or experiences. Observations compared to commercial software packages are listed in Table 2, which show the generality, efficiency and less required user background knowledge of our method.

The advantages of our method compared with other widely used methods in literature are as follows:

1.  Simple user interface: What the user just needs to do is to indicate a sparse point loop for the target region, import axes from preoperative planning results, and input related parameters. Compared with other traditional CAD software ([6,7,9,10]), it is very easy to learn and operate, and it allows the clinical user to design surgical template quickly and modify it conveniently.
2.  Generality: The method is applicable to template design for a variety of surgical interventions, while most of other methods in literature can only be employed for a specified surgery. For example, CoDiagnostiX[TM],

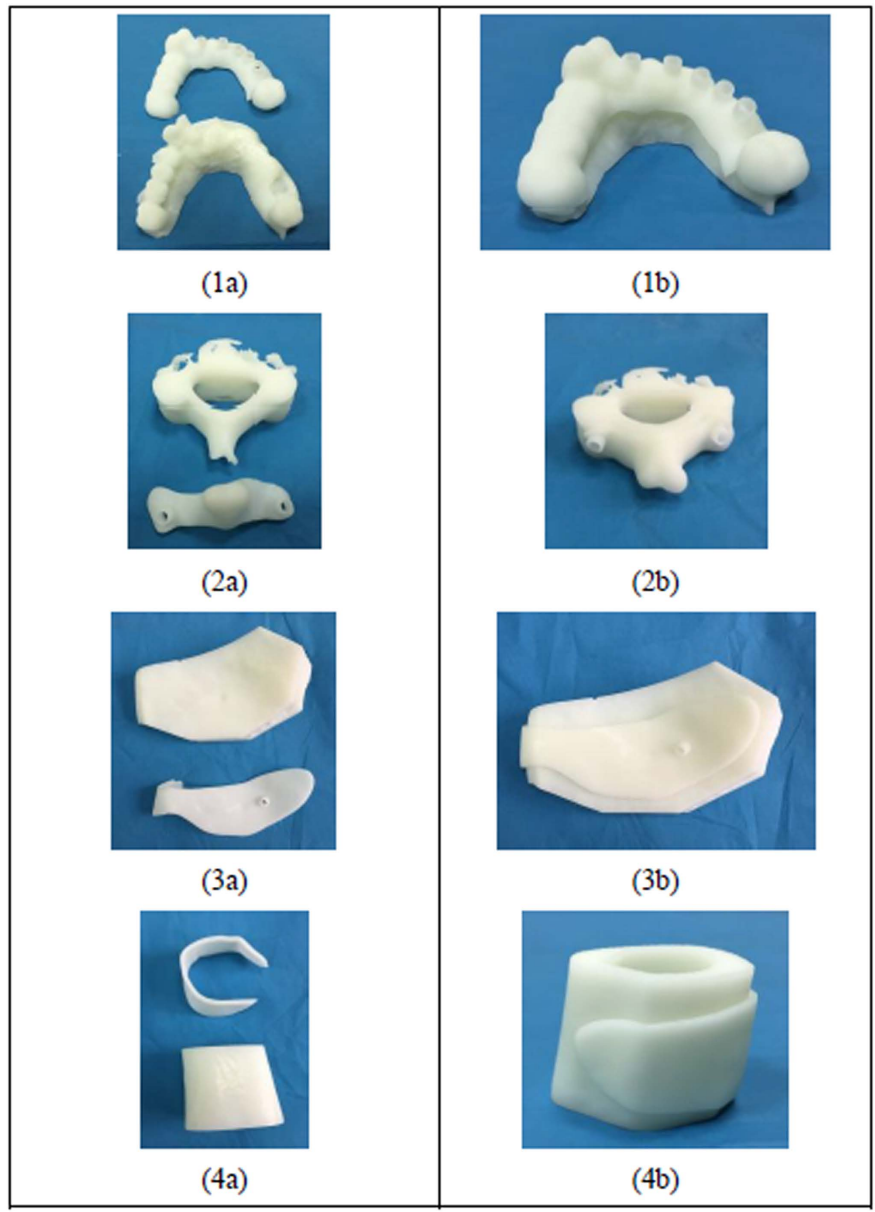

**Figure 7.** (**1a–4a**): The 3D-printed surgical templates and the adjacent tissue models (**1a**): mandibular phantom, (**2a**): part of cervical vertebrae phantom, (**3a**): part of cervical vertebrae phantom, (**4a**): part of bone phantom); (**1b–4b**): Matching of the surgical template with the adjacent tissue models.

3Shape Dental System™, and Nobel Biocare (NobelGuide), are just focusing on oral and maxillofacial surgery, while SPPC and Zimmer Patient Specific Instruments (Warsaw, IN, USA) are only applicable to TKA.

3. High efficiency: For an engineer familiar with traditional CAD software of mechanical design, it will cost several hours to design a template. Modification of template design according to the surgeon's demands will add another several days. As for the template designed and fabricated by the software supplier, the delivery time of several days is unavoidable. With our method, it only takes up to a few minutes to finish the template design, and modification is also very efficient. The concrete time cost depends on the mesh tessellation. The output model is in the form of STL so that it can be promptly fabricated through 3D printing technology.

However, a connected mesh is required for the segmentation algorithm in this study. Since "holes" may occur on the surface mesh reconstructed from CT or CBCT data, if those "holes" are quite close to the contour generated dynamically with the indicated control points, the generation of point loop for segmentation may fail, resulting in unexpected clipping result. Besides, the post-processing of the template is also required before it is applied to clinical practice. Similar to the template design using traditional CAD software of mechanical design, morphology of the inner surface is the inverse of the bone surface of the surgical site, guaranteeing unique fitness between the template and the surgical site. However, if the target region on the model surface is complex, for

| | | Oral implantology | Iliosacral screw insertion | | Cervical pedicle screw placement | | Osteotomy | |
|---|---|---|---|---|---|---|---|---|
| | Sampling step | 20 | 5 | 20 | 10 | 20 | 5 | 10 |
| **Number** | Triangles of input mesh | 213410 | 1073058 | 1073058 | 38082 | 38082 | 149054 | 149054 |
| | Points of input mesh | 106707 | 534619 | 534619 | 19146 | 19146 | 73813 | 73813 |
| | Edge points of inner surface | 1645 | 565 | 555 | 266 | 252 | 290 | 292 |
| | Triangles of inner surface | 41177 | 6967 | 6959 | 1240 | 1186 | 1300 | 1194 |
| | Points of inner surface | 21412 | 3767 | 3758 | 754 | 720 | 796 | 744 |
| | Drilling tubes | 5 | 1 | 1 | 2 | 2 | 0 | 0 |
| **Time(s)** | Inner surface segmentation | 7.956 | 13.837 | 12.643 | 0.796 | 0.671 | 2.028 | 3.214 |
| | Offset of inner surface | 11.793 | 1.576 | 1.545 | 1.357 | 1.341 | 1.778 | 1.513 |
| | Generation of points for outer surface segmentation | 7.301 | 5.787 | 2.075 | 2.683 | 2.34 | 6.614 | 3.728 |
| | Outer surface segmentation | 3.557 | 2.403 | 2.511 | 3.011 | 3.073 | 4.758 | 4.025 |
| | Connection of inner and outer surfaces | 0.609 | 0.203 | 0.172 | 0.141 | 0.14 | 0.187 | 0.156 |
| | Initial template generation | 31.216 | 23.806 | 18.946 | 7.988 | 7.565 | 15.365 | 12.636 |
| | Runtime of Boolean operation(s) | 16.723 | 2.621 | – | 4.977 | – | – | – |

**Table 1.  Scale of Input Models and the Runtime (Sampling step refers to the step of sampling inner surface edge points in progress of point generation for outer surface clipping).**

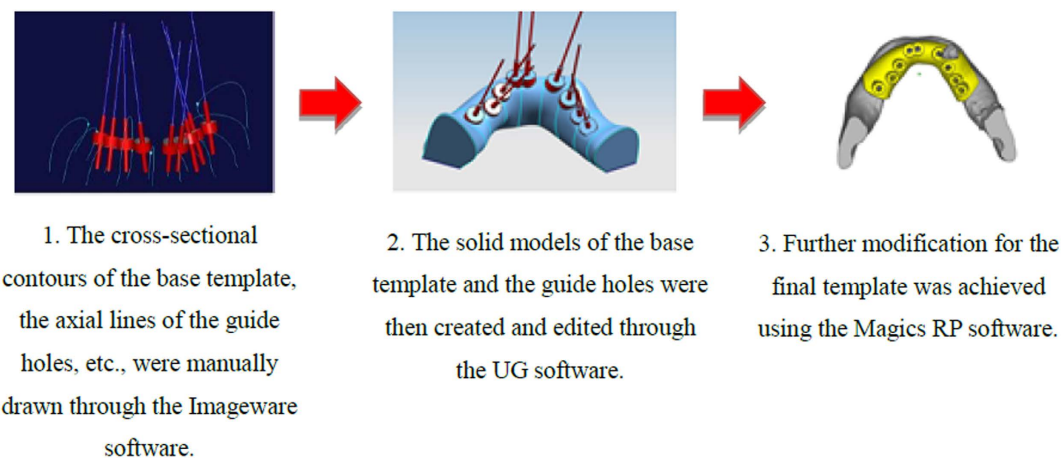

1. The cross-sectional contours of the base template, the axial lines of the guide holes, etc., were manually drawn through the Imageware software.

2. The solid models of the base template and the guide holes were then created and edited through the UG software.

3. Further modification for the final template was achieved using the Magics RP software.

**Figure 8.  The workflow of Method 1 for the template design.**

| | | Method 1: Using the Imageware, UG, and Magics RP together | Method 2: Using 3-matic (Materialise, Leuven, Belgium) | Method 3: Using TemDesigner |
|---|---|---|---|---|
| Oral implantology | Time(h) | 2-3 | 1-2 | 0.2-0.3 |
| | User interaction | Very Complicated | Complicated | Simple and Easy |
| | Required user experience | Very High | High | Low |
| Iliosacral screw insertion | Time(h) | 1-2 | 0.5-1 | 0.2-0.3 |
| | User interaction | Very Complicated | Complicated | Simple and Easy |
| | Required user experience | Very High | Medium | Low |
| Cervical pedicle screw placement | Time(h) | 2-3 | 1-2 | 0.2-0.3 |
| | User interaction | Very Complicated | Complicated | Simple and Easy |
| | Required user experience | Very High | High | Low |
| Osteotomy | Time(h) | 0.5-1 | 0.5-1 | 0.2 |
| | User interaction | Very Complicated | Complicated | Simple and Easy |
| | Required user experience | Very High | Medium | Low |

**Table 2.  Observations and comparison among commercial software packages and our proposed method.**

example, in some partially edentulous cases of oral implantology, "shortcut" may happen, i.e., bottom of the template is smaller than the upper part, and then it will not be assembled without additional tuning of the template inner surface. For further work, a segmentation algorithm that can deal with an unconnected mesh, will be developed. A feasible solution is to detect and patch the holes of the mesh before the segmentation. As for merging, the collision point loop will be used for clipping the two meshes separately. Then, points of the border edge for merging will be modified to ensure closure after merging. This software will be developed as a free extension module of 3D Slicer, the free, open source software package for visualization and image analysis (http://www.slicer.org/).

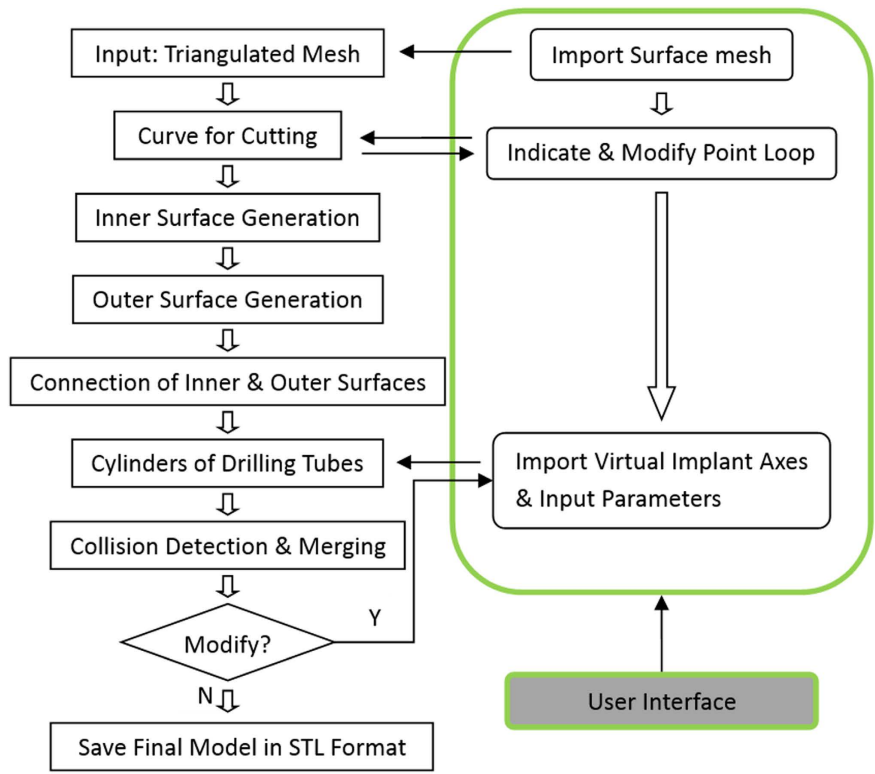

**Figure 9. General framework of semi-automatic surgical template design.**

In addition, the function of predicting the stability of the fit of a patient-specific template will be developed in the future work. In the previously published work, Van de Broeck *et al.*[34] have proposed a method to analyze the contact surface of a customized template and predict how stable the contact on the supporting bone surface will be. We plan to adopt their novel methods in the near future so that it will allow the surgeons to compare different designs during the preoperative design process.

## Methods

**Overview.**    The general framework of our approach is shown in Fig. 9 and the user interactions are described in details below:

1. Input Surface Mesh: Generally, 3D models reconstructed from CT or CBCT data are in STL format so that they are polyhedra with triangle faces. If not, the model will be triangulated firstly.
2. Curve for cutting: A closed curve on the surface mesh will be generated with the user points to indicate a target region. The curve can be modified dynamically through adjustment or cancel of the existing control points, or adding new ones.
3. Inner surface generation: With sample points from the curve, the target region is segmented from the input mesh as the inner surface of the template.
4. Outer surface generation: A closed offset surface of the inner surface is firstly achieved by means of distance field method. Then, mesh segmentation is utilized to clip the outer surface from the offset. Sampling points of the inner surface edges are projected to the offset surface to obtain points for clipping.
5. Connection of inner & outer surfaces: Inner and outer surfaces are connected with ruled surfaces. To address this problem, it is transferred to shortest path problem in a directed single layer graph.
6. Collision detection & merging: Firstly, collision detection is employed to generate the collision polylines and obtain intersection triangles. Secondly, intersection triangles are removed. Thirdly, extract the relevant parts and fuse them together with ruled surfaces.

**Inner surface generation.**    *Mesh segmentation.*    The inputs of the segmentation algorithm include: 1. A connected mesh; 2. A loop consisting of a not-self-intersecting point sequence and the points must be on the mesh. The general procedure is described as follows: Firstly, another loop strictly going along the edges (the line segments connecting adjacent vertices) of the mesh is created. Then, signed distances from each vertex to the initial loop are calculated. Finally, the signed distances are utilized as implicit function to clip the mesh.

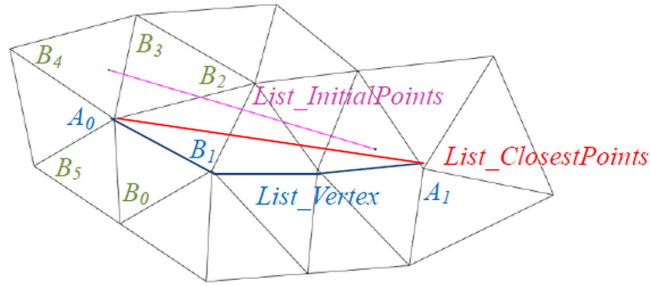

**Figure 10. Mesh edge tracking process: Among all neighbor vertices of** $A0$, **i.e.** $B_0$, $B_1$, $B_2$, $B_3$, $B_4$, $B_5$, **only** $B_0$, $B_1$, **and** $B_2$ **meet the condition** $< A_0A_1, A_0B_i > \in [0°, 90°]$. **Because** $B_1$ **is closest to** $A_0A_1$, **it is inserted to** *List_Vertex*. **Pink segment of** *List_InitialPoints*: **the initial segment of user placed points; Red segment of** *List_ClosestPoints*: **the segment of closest points; Blue polyline of** *List_Vertex*: **the polyline strictly going along the edges of the mesh.**

*Edge Loop Generation.*

1. Suppose *List_InitialPoints* denotes the initial, manually indicated point sequence. Connect adjacent points with linear segments in order, and the polyline is denoted as $L_{polyline}$.
2. Search the mesh vertex closest to each initial point and save the new vertex sequence as *List_Closest-Points* $= \{A_0, A_1, A_2 \ldots\}$.
3. The edges of the mesh connecting the closest vertices are tracked as the following procedure (shown in Fig. 10), and denote the new vertex sequence by *List_Vertex*, i.e. insert vertices between adjacent points in *List_ClosestPoints*, so that the polyline of the new vertex sequence *List_Vertex* is strictly along the edges of the mesh:

    i. Add $A_0$ to List_Vertex;
    ii. For current $Ai$, $P = Ai$;
    iii. Search all neighbor vertices of $P$, denoted $\{B_0, B_1, B_2, \ldots, B_n\}$;
    iv. Find the $B_i$ which meets the following condition, and insert it to *List_Vertex*;

$$< \overline{A_iA_{i+1}}, \overline{PB_i} > \in [0°, 90°] \; d_i \leq d_j (j = 0, 1, \ldots, n) \tag{1}$$

    in which $d_i$ is the distance from $B_i$ to line $A_iA_{i+1}$. That means, $B_i$ is oriented in the direction of and closest to vector $A_iA_{i+1}$.
    If $(B_i)$\{
        Insert $B_i$ to *List_Vertex*;
        If $(B_i \neq A_{i+1})$
            \{$P = B_i$; Go back to step iii);\}
        else
            \{$i = i + 1$; Go back to step ii);\}
        \}
        else\{Go to step v);\}
    v. Track the path between $A_i$ and $A_{i+1}$ along the edge using an approximate shortest path algorithm. Add the vertices to *List_Vertex*. Let $i = i + 1$ and go back to step ii;
    The algorithm terminates when $A_i$ becomes $A_0$ again.

**Calculation of signed distance.** Suppose the polyline $L_{polyline}$ of *List_InitialPoints* consists of $n$ segments, it also means there are totally $n$ points in the sequence. For every vertex of the mesh, distances to each of the $n$ segments are calculated and the smallest one is chosen as its scalar. Moreover, for each vertex $P_i$ in *List_Vertex*, the closest point to it lying within the segment which may finally contribute to the scalar is denoted as $Q_i$. After that, the sign of the distance is to be determined. Apparently, the loop of *List_Vertex* divides the mesh into two parts. The vertices belong to one part are marked positive, and the other negative. The signs of points in *List_Vertex* are determined as follows: 1. For $P_i$ in *List_Vertex*, among all its neighboring vertices of the mesh except those in *List_Vertex*, the one whose scalar is the largest is denoted as $N_i$; 2. If $\|N_iP_i\| < \|N_iQ_i\|$, the sign of $P_i$ is the same as $N_i$. Otherwise, its sign is opposite to $N_i$.

**Segmentation with signed distance.** Now, every vertex of the mesh has its own scalar, i.e. signed distance, which will be used to segment the surface. For each triangle of the mesh, three vertices are clockwise denoted as $P_0$, $P_1$, and $P_2$, respectively. If the scalar signs of $P_i$ and $P_{(i+1)\%3}$ are identical, then $P_i$ is saved as $Q_i$. If the signs of $P_0$, $P_1$, and $P_2$ are all positive or negative, $\Delta Q_0Q_1Q_2$ is the same triangle as $\Delta P_0P_1P_2$, and there is no new triangle generated. Otherwise, linear interpolation is utilized to insert new vertex, whose scalar is 0, on the edge $P_i$ $P_{(i+1)\%3}$ if the signs of $P_i$ and $P_{(i+1)\%3}$ are different. Then, a new triangle $\triangle Q_0Q_1Q_2$ is generated, dividing the original one $\triangle P_0P_1P_2$ into three. Figure 11 is an example demonstrating how the process works.

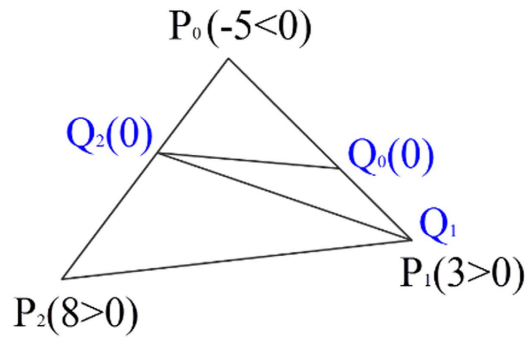

**Figure 11.  An example of clipping with signed distance: The scalars of $P_0$, $P_1$, $P_2$ are respectively $-5$, 3, 8. i)** $-5 \times 3 < 0$. Then, $Q_0$ is linearly interpolated at zero within segment $P_0P_1$. **ii)** $3 \times 8 > 0$. So, $Q_1$ is the same as $P_1$. **iii)** $8 \times (-5) < 0$, so $Q_2$ is linearly interpolated at zero. The new triangle $\triangle Q_0Q_1Q_2$ divides the original one $\triangle P_0P_1P_2$ into three triangles.

**Points for segmentation.**    Initial control points are placed by the user to surround a target region on the three-dimensional model surface. These control points can be modified, deleted, or added to adjust the target region. Usually, the initial points won't be dense enough for a segmentation result with smooth edge, taking into account the user-friendliness. Therefore, a method of resampling the spline generated with initial control points more densely was also developed in this study. First of all, cardinal spline is applied for interpolating. However, although the spline passes through the initial points on the surface, the spline itself is not located on the surface. That means, the interpolating points for clipping are most likely not on the surface. To solve this problem, at each current angle of view, world point coordinates of the resample points are converted to display coordinates, which will be converted to world point coordinates again. In consideration of efficiency, the z-buffer method is employed to carry out the second conversion instead of geometric methods, typically a ray cast. The conversion operation is updated in *real-time* when the next initial point is placed or the last one is deleted. So, if the angle of view is changed, current resample points after coordinate conversion will be saved. After that, the next initial control point placed by the user will start a new spline snippet at the new view angle. The two snippets of two different view angles are connected with smooth curve. Figure 12 demonstrates the curve generated with user defined points and the result of mesh segmentation.

**Outer surface generation.**    As explained above, distance field is contoured to generate isosurface instead of triangle or vertex offset directly. Firstly, a closed isosurface of the inner surface is generated. Then, the mesh segmentation algorithm described above is utilized. The closed surface will be clipped with the points projected from the inner surface border edges to generate the outer surface.

**Offset surface.**    Firstly, the distance field of the inner surface is established with the method of Payne and Toga[35]. A cuboid region, expanded in each of the x-y-z directions properly from the bounding box of the inner surface, is defined. The minimum distance of each sample point in the cuboid region to the inner surface is calculated with the octree-based spatial search method. Subsequently, the distance field is contoured to generate isosurface with specified value of thickness of the template. Then, the method of marching cubes is utilized. Each time eight points of a volume are processed by comparing their scalar value with the specified contour value. Then, through linear interpolating, the vertices of new polygons are obtained. The individual polygons passing through the cubes are finally fused into the isosurface. Figure 13 shows an example of the surface offset result.

**Clipping to generate outer surface.**    The mesh segmentation algorithm is utilized to obtain the outer surface of template. So, point loop on the mesh for clipping is necessary. Since the outline of the outer surface should be similar to the inner one, sampling points of inner surface border edges are projected to the closed surface to obtain segmentation points. To avoid self-intersection of the point loop, for each sampling point, the projection is achieved along the average normal direction within several neighbor points. After clipping is done, the part along with inner surface normal is chosen as the outer surface.

**Ruled surface.**    To ensure that the model is closed, the border edge points of the connection surface must be just those of the inner and outer surfaces. In order to simplify the algorithm and reduce time costing, ruled surface, i.e., sequence of triangles is utilized to connect inner and outer surfaces. Here, determination of the ruled surface is formulated as some certain kind of shortest path problem in a directed single layer graph[29]. Then label setting algorithm[36] is employed to solve the unconstrained single-source shortest path problem.
   Suppose the edge contours of inner and outer surfaces, which are actually polylines, be defined by the sequences of $m$ and $n$ ordered discrete points separately. That is, sequence of $P_0$, $P_1$, $P_2$, …, $P_{m-1}$ represents p, the border edge contour of inner surface, and sequence of $Q_0$, $Q_1$, $Q_2$, …, $Q_{n-1}$ represents $q$, the outer one. It is noted, that $P_0$ follows $P_{m-1}$, and $Q_0$ follows $Q_{n-1}$, so $p$ and $q$ separately consists of $m$ and $n$ segments. Here, an approximately optimized result can meet the command so that the two contours are handled as open for reducing complexity. Since one triangle consists of three vertices, it must be $\{P_i, P_j, Q_k\}$ or $\{Q_i, Q_j, P_k\}$. Thus, each triangle consists of one contour segment and two spans. Take $\{P_i, P_j, Q_k\}$ for example, one contour segment is $P_iP_j$, two spans are $P_iQ_k$ and $P_jQ_k$. To ensure that all the m + n points of the contours are the vertices of the ruled surface while avoiding self-intersection, two conditions should be met:

1. For two spans $P_iQ_j$ and $P_kQ_l$ either $i \leq k$ & $j \leq l$, or $i \geq k$ & $j \geq l$ is met;
2. The contour segment should be formed with two adjacent vertices.

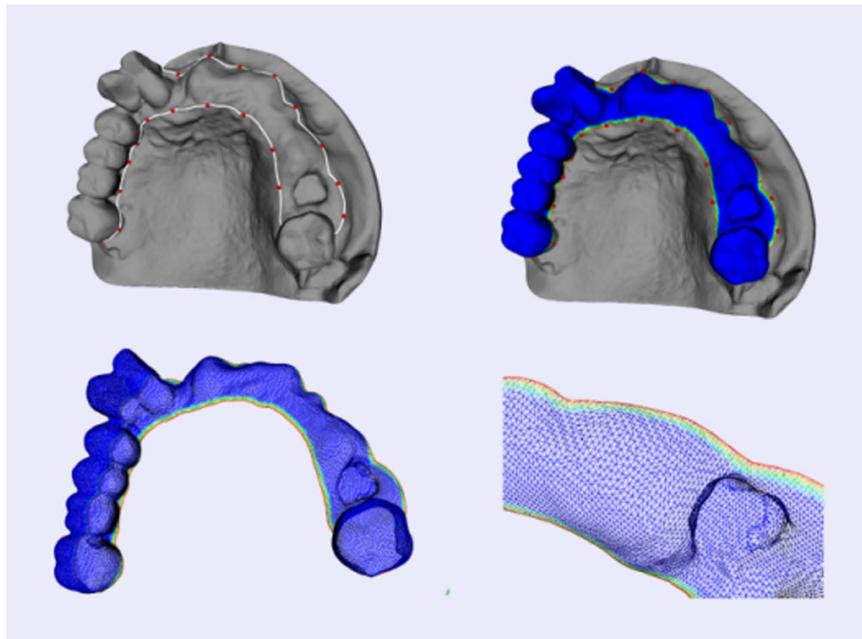

**Figure 12. Mesh segmentation: The white curve is generated with user points to indicate a target region.**

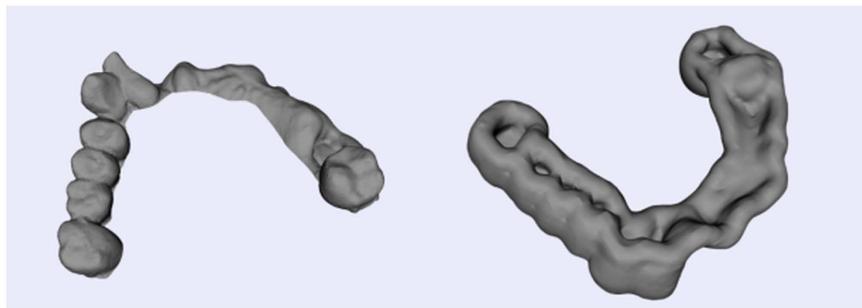

**Figure 13. Offset surface generation: the inner surface of an oral implantology template (left) and the offset surface of the inner surface (right).**

So, each triangle is defined as either of the form $\{P_i, P_{i+1}, Q_k\}$ or $\{Q_i, Q_{i+1}, P_k\}$, and each contour segment only contributes to one triangle. Obviously, there are totally m + n − 2 contour segments so that the ruled surface consists of $m + n - 2$ triangles. Taking into consideration of the two conditions above, for the triangle $\{P_i, P_{i+1}, Q_k\}$, its adjacent triangle with the common span $P_{i+1}Q_k$ can only be defined as the form of $\{P_{i+1}, P_{i+2}, Q_k\}$ or $\{P_{i+1}, Q_k, Q_{k+1}\}$. The next triangle can only be $P$-succeeded or $Q$-succeeded. In other words, if the next span is determined, the next triangle is determined. Then the determination of ruled surface is turned into the determination of the sequence of spans. It is easy to point out that the possible sequences which satisfy the criteria are totally

$$\binom{m+n-2}{m-1} = \binom{m+n-2}{n-1} = \frac{(m+n-2)!}{(m-1)!\,(n-1)!} \tag{2}$$

Graph theory is employed to calculate the shortest path for the selection of the proper sequence, whose sum of span length is the smallest.

The directed single layer weighted graph *G-(V, A)* is constructed as follows (shown in Fig. 14):

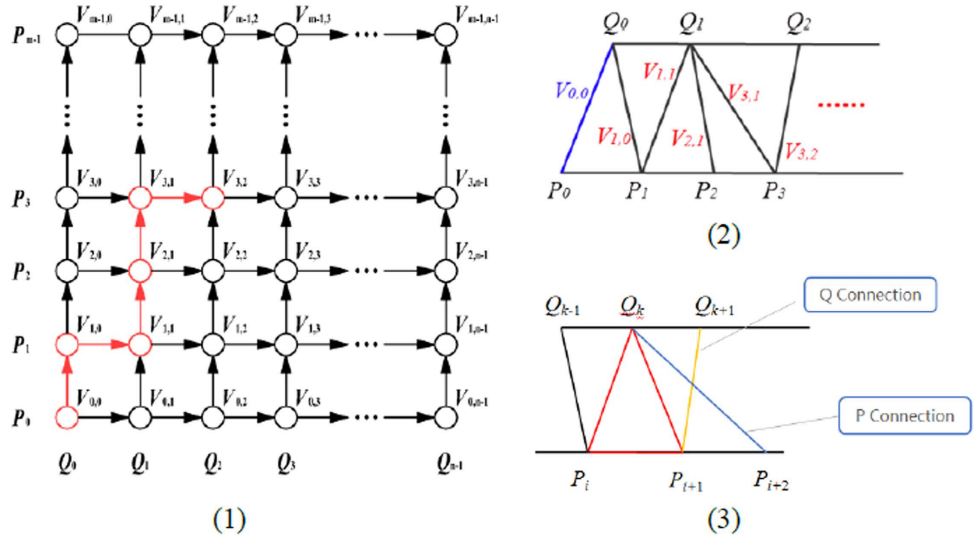

**Figure 14.** (**1**) Directed single layer weighted graph $G$-$(V, A)$. (**2,3**) The triangle organization of ruled surface corresponds to the red nodes in the first one.

1. Each node $V_i$, $\epsilon V$ in Row $i$, Column $j$ ($i = 0, 1, 2, …, m - 1; j = 0, 1, 2, …, n - 1$) represents a corresponding span $P_iQ_j$;
2. Each node $V_{i,j}$ is only incident to $V_{i+1,j}$ and $V_{i,j+1}$, and incident from $V_{i-1,j}$ and $V_{i,j-1}$. As is shown in the figure, each arc in set $A$ connect two adjacent nodes, oriented upward or to right;
3. The weight of the arcs incident to $V_{i,j}$ is the length of span $P_iQ_j$;

A link edge in the graph indicates a unique triangle whose left span is represented by the start node of the link edge, and right span corresponding to the end node. Thus, any path from the source node $V_{0,0}$ to the target node $V_{m-1,n-1}$ indicates a unique ordered sequence of triangle. Consequently, the question is reduced to a single-source shortest path problem which is then solved with label setting algorithm.

In this case, the label of a node $V_{i,j}$ is denoted by ( $i, j, length, Dis_{i,j}, Prev$ ), where $i$ and $j$ are subscripts of the node, $length$ means the weight of arcs incident to $V_{i,j}$, $Dis_{i,j}$ represents the shortest distance from the source node $V_{0,0}$ to the current node $V_{i,j}$, and $Prev$ indicates in the shortest distance case, the node before $Vi.j$ is on its left (denoted by 1) or downstairs ($-1$), since

$$Dis_{i,j} = \min\left(Dis_{i-1,j}, Dis_{i,j-1}\right) + length \tag{3}$$

We only need to compare $Dis_{i-1,j}$ and $Dis_{i,j-1}$ when the label of a node $V_{i,j}$ is to be updated.
The general flow of the label setting algorithm works as below:

1. Labels of nodes in $Row0$ and $Column0$ are updated. That's because in any path their previous node can only be the left and lower one respectively. $Dis_{i,j}$ of $V_{0,0}$ is set 0.
2. Labels of the rest nodes are calculated according to the index from small to large order as:

for i = 1:m − 1

    for $j = 1:n - 1$
      if ($Dis_{i-1,j} < Dis_{i,j-1}$)
        $Dis_{i,j} = Dis_{i-1,j} + lengh; Prev = -1$
         else
        $Dis_{i,j} = Dis_{i,j-1} + lengh; Prev = 1$
      for ends
    for ends

After the label of the target node $V_{m-1, n-1}$ is updated, the shortest path can be derived easily on the basis of $Prev$ in the labels from the target node to the source, and Fig. 15 shows the result of automatic connection of inner and outer surfaces of oral implantology template using the above algorithm.

**Collision detection and merging.** After connection of the inner and the outer surfaces, the initial template is generated. Then, drilling tubes need to be added. Firstly, collision detection is executed to generate the collision polylines and obtain intersection triangles. Secondly, intersection triangles are removed. Thirdly, extract the relevant parts and merge them together through collision polylines with ruled surfaces described above.

**Collision detection.**

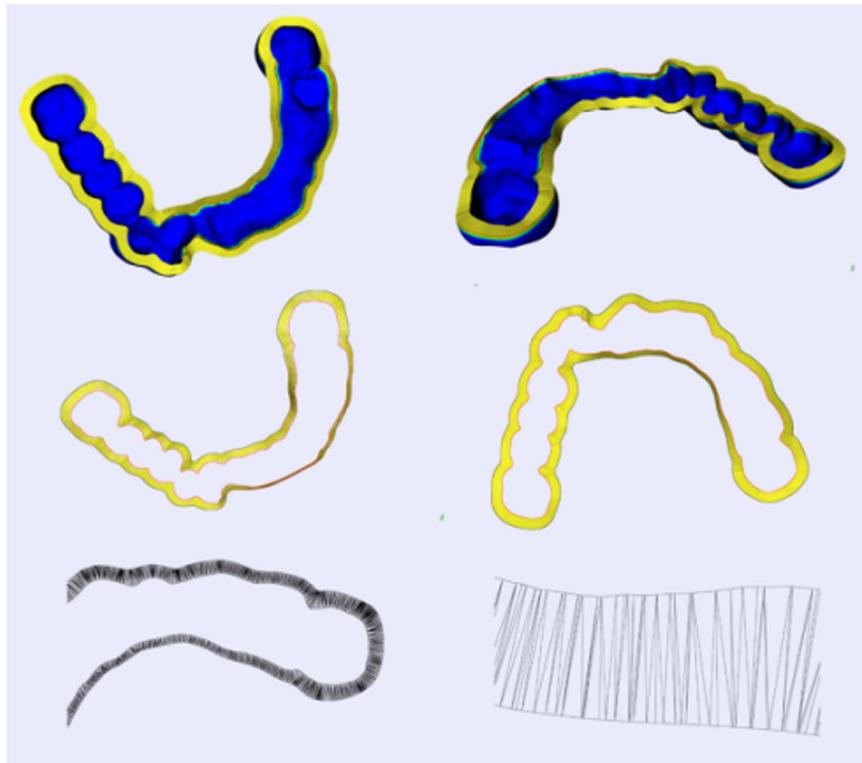

**Figure 15. Automatic connection of inner and outer surfaces of oral implantology template.**

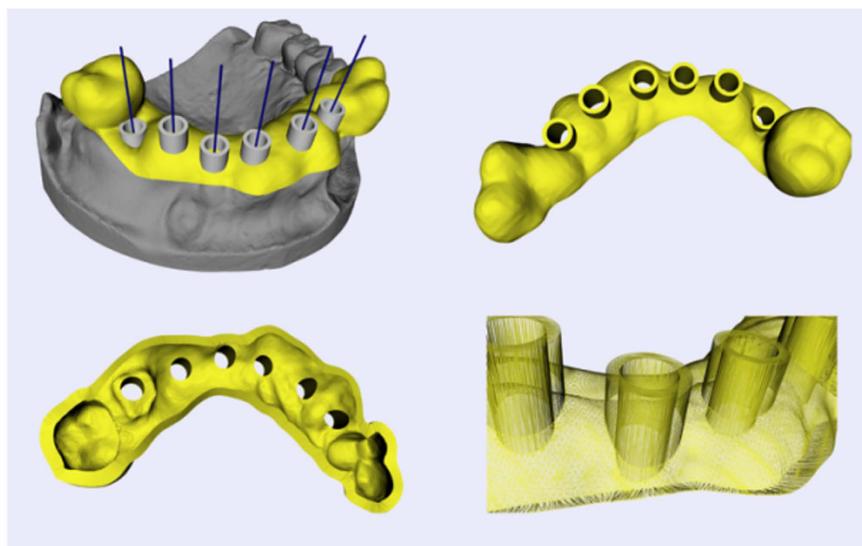

**Figure 16. Boolean operation of initial template for oral implantology and drilling tubes.**

1. Oriented bounding box (OBB) trees of the two triangulated surfaces for collision detection are generated.
2. Each pair of intersecting leaf nodes is detected.
3. Each triangle of the first node will be extracted sequentially to perform collision detection with all the triangles of the other node one by one. The two triangles of a pair are denoted Triangle1 and Triangle2 separately. Concrete procedure goes as follows:

    *for each edge of Triangle1*
      *intersection detection with the bounding box of Triangle2*
        *if intersects*
          *intersection detection with the finite plane of Triangle2*

*if intersects and the intersection point is inside Triangle2*
*the intersection point will be saved and the two triangles be marked*
*else if the two triangles are coplanar and overlapping partially*
*the intersection point of two edges is saved*

**Merging.** After the collision detection is finished, one or more polylines are generated by connecting the intersection points together in order. The polyline can be used as point loop for clipping the two surfaces separately. Then, relevant parts are extracted to be fused together. However, different geometric structures, mainly constructions of triangles of the two surfaces, lead to different results of clipping edge, so the two parts can't match exactly at the clipping edge without additional process. In our case, the polyline of intersection points is utilized as the joint edge. All the marked triangles are removed and each original triangulated surface is divided to at least two fragments. Finally, the target parts of each original surface are merged together through filling triangles between the correct edge of fragment and the polyline (as shown in Fig. 16).

## Acknowledgements

This study was supported by Natural Science Foundation of China (81171429, 81511130089), Shanghai Pujiang Talent Program (13PJD018), and Foundation of Science and Technology Commission of Shanghai Municipality (14441901002, 15510722200). Dr. Dr. Jan Egger receives funding from the European grants ClinicIMPPACT (610886) and GoSmart (600641), and BioTechMed-Graz ("Hardware accelerated intelligent medical imaging").

## Author Contributions

X.C. and L.X. conceived of the project, and designed the framework of the software. X.C., L.X. and Y.Y. proposed the new algorithms, developed the software and performed the experiments. X.C., L.X., Y.Y. and J.E. wrote the paper. All authors discussed the results and commented on the manuscript.

## Additional Information

**Competing financial interests:** The authors declare no competing financial interests.

**How to cite this article**: Chen, X. *et al.* A semi-automatic computer-aided method for surgical template design. *Sci. Rep.* **6**, 20280; doi: 10.1038/srep20280 (2016).